\pdfoutput=1
\documentclass[conference]{IEEEtran}

\usepackage{graphicx}
\usepackage{tabularx}

\usepackage[T1]{f ontenc}
\usepackage[utf8]{inputenc}

\usepackage{csquotes}
\usepackage{amsthm}
\usepackage{nicefrac}
\usepackage[textsize=tiny]{todonotes}
\usepackage{booktabs}
\usepackage{xcolor}
\usepackage{tikz}
\usetikzlibrary{arrows,positioning} 
\usetikzlibrary{shapes.geometric, positioning, quantikz}
\usetikzlibrary{positioning}
\usepackage{nicematrix}
\usepackage{flushend}
\usepackage{hyperref}
\usepackage[binary-units=true, detect-all=true]{siunitx}
\usepackage{etoolbox} 
\usepackage{soul}
\usepackage{amssymb}

\robustify\bfseries

\makeatletter
\let\MYcaption\@makecaption
\makeatother
\usepackage[font=footnotesize]{subcaption}
\makeatletter
\let\@makecaption\MYcaption
\makeatother

\usepackage{yquant}
\usepackage{pgfplotstable}

\usepackage[style=ieee, maxnames=3, minnames=1, date=year, url=false, doi=false,isbn=false,backend=biber]{biblatex}
\AtEveryBibitem{%
  \clearname{editor}%
  \clearfield{series}%
  \clearfield{isbn}%
  \clearfield{issn}%
    \clearfield{volume}%
  \clearfield{number}%
  \clearfield{pages}%
  \ifentrytype{misc}{}{\clearfield{eprint}}
}

\def\circlesize{2cm}
\def\horizontalspace{-6mm}

\newtheorem{example}{Example}

\definecolor{color1}{RGB}{33, 145, 140}
\definecolor{color2}{RGB}{64,1,84}

\begin{document}

\title{Predicting Good Quantum \\Circuit Compilation Options}

\author{
	\IEEEauthorblockN{Nils Quetschlich\IEEEauthorrefmark{1}\hspace*{1.5cm}Lukas Burgholzer\IEEEauthorrefmark{2}\hspace*{1.5cm}Robert Wille\IEEEauthorrefmark{1}\IEEEauthorrefmark{3}}
	\IEEEauthorblockA{\IEEEauthorrefmark{1}Chair for Design Automation, Technical University of Munich, Germany}
	\IEEEauthorblockA{\IEEEauthorrefmark{2}Institute for Integrated Circuits, Johannes Kepler University Linz, Austria}
	\IEEEauthorblockA{\IEEEauthorrefmark{3}Software Competence Center Hagenberg GmbH (SCCH), Austria}
	\IEEEauthorblockA{\href{mailto:nils.quetschlich@tum.de}{nils.quetschlich@tum.de}\hspace{1.5cm}\href{mailto:lukas.burgholzer@jku.at}{lukas.burgholzer@jku.at}\hspace{1.5cm} \href{mailto:robert.wille@tum.de}{robert.wille@tum.de}\\
	\url{https://www.cda.cit.tum.de/research/quantum}
	}
}

\maketitle

\begin{abstract}
Any potential application of quantum computing, once encoded as a quantum circuit, needs to be compiled in order to be executed on a quantum computer.
Deciding which qubit technology, which device, which compiler, and which corresponding settings are best for the considered problem---according to a measure of \emph{goodness}---requires expert knowledge and is overwhelming for \mbox{end-users} from different domains trying to use quantum computing to their advantage.
In this work, we treat the problem as a statistical classification task and explore the utilization of supervised machine learning techniques to optimize the compilation of quantum circuits.
Based on that, we propose a framework that, given a quantum circuit, predicts the best combination of these options and, therefore, \emph{automatically} makes these decisions for \mbox{end-users}.  
Experimental evaluations show that, considering a prototypical setting with $3000$ quantum circuits, the proposed framework yields promising results: for more than three quarters of all unseen test circuits, the \emph{best} combination of compilation options is determined. 
Moreover, for more than $95\%$ of the circuits, a combination of compilation options within the \mbox{top-three} is determined---while the median compilation time is reduced by more than one order of magnitude.
Furthermore, the resulting methodology not only provides \mbox{end-users} with a prediction of the best compilation options, but also provides means to extract explicit knowledge from the machine learning technique.
This knowledge helps in two ways: it lays the foundation for further applications of machine learning in this domain and, also, allows one to quickly verify whether a machine learning algorithm is reasonably trained.
The corresponding framework and the \mbox{pre-trained} classifier are publicly available on GitHub (\url{https://github.com/cda-tum/MQTPredictor}) as part of the Munich Quantum Toolkit (MQT).
\end{abstract}

\section{Introduction}
The capabilities of quantum computers are steadily evolving---achieving more physical qubits, lower error rates, and faster operations.
Devices are increasingly made available through manufacturers, such as IBM Quantum, or third-party cloud providers, such as Amazon Web Services.
Furthermore, several \emph{Software Development Kits} (SDKs) for programming
these devices have been developed, e.g., IBM's Qiskit~\cite{qiskit}, Quantinuum's TKET~\cite{sivarajahKetRetargetableCompiler2020}, Google's Cirq~\cite{cirq}, Xanadu's Pennylane~\cite{bergholmPennyLaneAutomaticDifferentiation2020}, and Rigetti's Forest~\cite{rigetti}.
As a result of this progress, academia and industry have started to elaborate possible applications for this new technology. 
In fact, the potential of quantum computing is currently being explored for several application domains such as chemistry (e.g.,~\cite{peruzzoVariationalEigenvalueSolver2014}), finance (e.g.,  ~\cite{stamatopoulosOptionPricingUsing2020}), optimization (e.g.,~\cite{harwoodFormulatingSolvingRouting2021}), and machine learning (e.g.,~\cite{zoufalQuantumGenerativeAdversarial2019}).
Even dedicated workflows describing the steps required to solve a classical problem using quantum computing start to emerge (e.g.,~\cite{quetschlich2023mqtproblemsolver}).

The process towards executing corresponding quantum algorithms on quantum computers has a lot of similarities to the process of executing classical algorithms or programs on classical computers.
In the classical world, a program must be transformed so that it can be executed on a specific \emph{Central Processing Unit}~(CPU) which provides its own \emph{Instruction Set Architecture}~(ISA).
This transformation process is called \emph{compilation} and is conducted by \emph{compilers}---usually coming with numerous settings and optimization parameters. 
Similarly, once an application for quantum computing has been encoded as a quantum circuit, it needs to be \emph{compiled} to the targeted device's native gate-set while obeying all restrictions imposed by the device, e.g., limitations on the interaction of qubits.

For classical compilation, guidelines and \mbox{best-practices} have been developed over the previous decades such that even \mbox{end-users} without a background in computer science can compile and execute their programs.
In quantum compilation, however, this is not yet the case and classical compilation schemes cannot be adopted in a \mbox{one-to-one} fashion.
The reason for that lies within the constraints (such as the restricted connectivity of quantum devices) to be satisfied by a compiled quantum circuit compared to classical software.
Since comparable guidelines and \mbox{best-practices} have not yet been developed for quantum circuit compilation, expert knowledge is required and especially \mbox{end-users} without a background in quantum computing are easily overwhelmed with a flood of different qubit technologies, devices, compilers, and (quite frequently, poorly documented) settings to choose from.
Without actionable advice, it is extremely difficult for an \mbox{end-user} to decide on a combination of all these options for a correspondingly considered application.

In this work, we treat the problem as a statistical classification task and explore the utilization of supervised machine learning techniques to optimize the compilation of quantum circuits.
We propose a framework that, given a quantum circuit, predicts the best combination of these options and, by that, \emph{automatically} decides for \mbox{end-users} which qubit technology, which specific device, which compiler, and which corresponding settings to choose for realizing their applications.
By this, a similar kind of comfort is provided to the \mbox{end-users} of quantum computers as is taken for granted in the classical world.

\vspace{1cm}
In order to demonstrate the effectiveness of the proposed methodology, we trained a machine learning classifier on $3000$ training circuits (taken from the MQT Bench library~\cite{quetschlich2022mqtbench}) considering two qubit technologies, five different devices, two compilers, and six corresponding settings.
The resulting classifier, which is publicly available along with the proposed framework, is shown to yield the best possible combination of compilation options in more than three quarters of all unseen test circuits. 
For more than $95\%$ of the circuits, a combination of compilation options within the \mbox{top-three} is determined---with the median compilation time being reduced by more than one order of magnitude compared to manually compiling for all possible combinations of compilation options and choosing the best result.
Moreover, rather than functioning as a \mbox{black-box}, the underlying model additionally provides means to extract explicit knowledge from the classifier---potentially guiding further works towards exploiting the full potential of machine learning in this domain and, also, to quickly verify whether a machine learning algorithm is reasonably trained.
The corresponding framework and the \mbox{pre-trained} classifier are publicly available on GitHub (\url{https://github.com/cda-tum/MQTPredictor}) as part of the Munich Quantum Toolkit (MQT).

The rest of this work is structured as follows:
In \autoref{sec:considered_problem}, we describe the process of determining a good
combination of compilation options and review the related work. 
Based on that, \autoref{sec:mqtpredictor} details the methodology to predict good combinations of options.
\autoref{sec:prob_desc} describes the resulting framework and summarizes our experimental evaluations.
\autoref{sec:discussion} discusses how explicit knowledge can be extracted from the proposed methodology and how changes in the underlying data can be incorporated before \autoref{sec:conclusion} concludes this work.

\section{Considered Problem: \\Determining Good Compilation Options}
\label{sec:considered_problem}
In this section, we review the spectrum of options to choose from when realizing an application on an actual quantum computer, discuss the resulting dilemma for \mbox{end-users}, and the related work available on this topic so far.

\subsection{Motivation}
\label{sec:motivation}

Due to its promising applications and the steady evolution of the corresponding technologies, quantum computing is no longer a niche topic.
Researchers in application domains very different from quantum computing seek to utilize this technology as a tool to solve their \mbox{domain-specific} problems in a more efficient fashion.
As a consequence, quantum computers are no longer exclusively used by physicists or computer scientists, but by an interdisciplinary range of \mbox{end-users}.

After a quantum algorithm for solving a particular problem from an application domain has been developed in terms of a quantum circuit, the \mbox{end-user} is confronted with a flood of possible options and design decisions:
\begin{itemize}
	\item Which \emph{qubit technology}, e.g., superconducting qubits or ion traps, is best suited for the application at hand?
	\item Which particular \emph{device}, e.g., from IBM or Rigetti, fits the quantum algorithm best?
	\item Which \emph{compiler}, such as IBM's Qiskit or Quantinuum's TKET, is most efficient for compiling the algorithm to the respective device?
	\item Which \emph{settings and optimizations} offered by the compilers, such as different levels of optimization, are adequate for the problem considered?
\end{itemize}

Figuring out good combinations of options (according to some evaluation metric) for a particular application is a highly \mbox{non-trivial} task due to several factors:
\begin{itemize}
	\item Different qubit technologies 
	have their own advantages and disadvantages, e.g., devices based on \mbox{ion-traps} have an all-to-all connectivity but suffer from slow gate execution times, while devices based on superconducting qubits have limited qubit connectivity but fast gate execution times~\cite{comparisontechnologies}.
	\item Individual devices greatly vary in their characteristics such as qubit count, error rates, coherence times, qubit connectivity, and gate execution times.
	\item Various compilers have been proposed by industry and academia, i.e., in \cite{qiskit,sivarajahKetRetargetableCompiler2020, rigetti, bergholmPennyLaneAutomaticDifferentiation2020, cirq, compiler1, compiler2, compiler3}---each of which is particularly suited for certain classes of circuits and architectures.
	\item Compiler settings and optimizations, quite frequently, are hardly or insufficiently documented.
	\item On top of all that, the domain of quantum computing is fast-paced and constantly changing---quickly and frequently redefining the state of the art.
\end{itemize}

This leaves \mbox{end-users} without expert knowledge faced with more options than they can feasibly explore.
Since quantum computing is still in its infancy, this diversity of options is only going to increase in the future.
Thus, automated tools for predicting good combinations of options are absolutely necessary in order to provide the same kind of comfort that is taken for granted in the classical world today.
Otherwise, we might end up in a situation where we have powerful quantum computers and tools, but only a selected group of people know how to use them.

\subsection{Related Work}\label{sec:related}
Decades of work on classical compilers have led to many sophisticated compiler optimization techniques.
\emph{Autotuning} (which describes the process of trying different compilation parameters and comparing their result against some metric) and machine learning (overviews are given in, e.g., \cite{ml_in_compilers, autotuning_compiler}) have shown promising results and established themselves as \mbox{state-of-the-art} techniques in this domain.
Even combined approaches of those two techniques are explored in~\cite{agakov_comb}, where a machine learning model is used not to directly determine optimal compiler settings but to identify promising areas of the optimization space. 
Additionally, reinforcement learning gained a lot of attention in recent years with promising results, e.g., in \cite{reinforcement1,reinforcement2,reinforcement3}---all targeting the LLVM~\cite{llvm} compiler.

In quantum compilation, a domain that is still in its infancy, the applied techniques are not that highly developed.
Nevertheless, first works towards this direction have already been proposed nearly a decade ago in the form of hardware resource estimates to reliably execute various quantum algorithms~\cite{suchara_qure_2013} to provide guidance to \mbox{end-users}.
Until today, resource estimation has still been an open research question. 
Microsoft recently proposed the Azure Quantum Resource Estimator~\cite{azureressourceestimator2022} which is publicly available in their SDK.
Furthermore, a \mbox{application-specific} resource estimator in the domain of derivative pricing has been developed~\cite{Chakrabarti2021thresholdquantum} and can be used to estimate the minimal quantum computing resources needed to achieve a \mbox{real-world} quantum advantage in this domain.
However, resource estimation often requires \mbox{end-users} to manually provide a lot of information on particular hardware aspects such as gate times and fidelities.

A different, but related, approach is to evaluate whether existing architectures and compilers are capable of executing a given quantum circuit.
One example is the \emph{NISQ Analyzer} proposed in \cite{salm_nisq_2020} which allows one to determine architectures that are expected to be capable of reliably executing a given circuit based on their number of qubits and a measure of their maximum supported circuit depth.
Although this solution requires less manual input, it still provides no advice on which compiler (settings) to use and also on which of the determined architectures to pick.

In the recent past, many tools for comparing the performance of different quantum circuit compilers for different devices and/or compilation options have been \mbox{proposed~\cite{salm_automating_2021, ArlineQuantumAppliedMachine, millsApplicationMotivatedHolisticBenchmarking2021, lubinskiApplicationOrientedPerformanceBenchmarks2021a}}.
However, in all these approaches, each input circuit is simply compiled for every possible combination of compilation options and the best circuit according to some metric is reported,~i.e., the best option is determined in a \mbox{brute-force} fashion.
While this provides the basis for interesting case studies, such an approach cannot be feasibly employed in practical situations due to the sheer amount of time it takes to try out all options on every invocation with a particular circuit---a situation which is only going to get worse with the ongoing increase in options.

In addition to the above approaches, machine learning methods have been proposed to tackle certain quantum compilation challenges by learning from data.
In~\cite{salmPrio2022}, the influence of several circuit characteristics on the quality of the circuit execution results is studied using \emph{\mbox{Multi-Criteria} Decision Analysis} methods and machine learning approaches to optimize such.
Similarly, the approach proposed in~\cite{wangQuEestGraphTransformer2022} tries to learn an estimate of circuit fidelity for specific devices using the graph structure of quantum circuits.
Complementarily, a reinforcement \mbox{learning-based} approach that models quantum compilation as a \emph{Markov Decision Process} with the goal of learning optimal compilation sequences has recently been proposed in~\cite{quetschlich2023compileroptimization}.

Overall, there are many interesting activities happening in this area.
Nevertheless, to the best of the authors' knowledge, there is no automatic solution to determine the best combination of quantum compilation options without explicitly executing and examining all of them---which might be feasible for conducting a case study on a handful of devices and compilers but is certainly not scalable enough to support \mbox{end-users} in realizing their \mbox{real-world} applications. 
In this work, we apply similar techniques as in classical compiler optimization targeting this problem.

\vspace{2cm}
\section{Proposed Solution: \\Predicting Good Compilation Options}
\label{sec:mqtpredictor}
As discussed in the previous section, naively executing and evaluating all possible combinations of compilation options on demand quickly becomes infeasible due to the large number of possible options.
Even if an \mbox{end-user} was able to utilize all those different SDKs to compile the specific quantum circuit for all computers, it would have been a very time-consuming and costly endeavor. 
Thus, in this work, we propose a methodology that allows one to \emph{predict} good combinations of options without the need for explicit compilation.
To this end, the problem is interpreted as a statistical \emph{classification} task which constitutes a prime task for \emph{supervised machine learning} algorithms.
In the following, each individual aspect of the proposed methodology is described and illustrated exemplary. Based on that, \autoref{sec:prob_desc} then covers how this results in a corresponding framework for the prediction of good combinations of options and evaluates the benefits for \mbox{end-users}.

\subsection{Compilation Options and Compilation Pipeline}\label{sec:pipeline}

Given certain qubit technologies with a set of respective devices and certain compilers with various settings, the search space of all compilation options is spanned by all the possible combinations of technologies, devices, compilers, and settings.
Therefore, a compilation pipeline needs to be set up to realize each combination of compilation options as a basis for the proposed methodology.
This poses a significant challenge as a result of the vastly different interfaces and usability levels of existing SDKs.
Overall, this results in a decision tree structure, where each path from the root node to a leaf node represents one possible combination of compilation options.

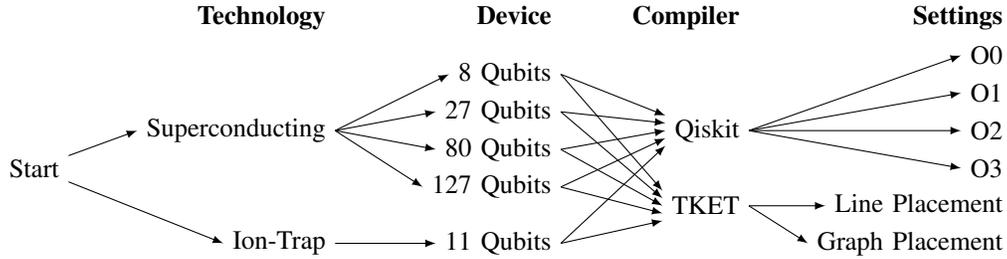
\begin{figure*}
\centering
\begin{tikzpicture}
  \node (start) at (-5,0.75)   {Start};
  
  \node[anchor=east] at (-1,2.75)   {\textbf{Technology}\strut};
  \node[anchor=east] at (2,2.75)   {\textbf{Device}\strut};
  \node[anchor=east] at (4.5,2.75)   {\textbf{Compiler}\strut};
  \node[anchor=east] at (8,2.75)   {\textbf{Settings}\strut};
  
  \node[anchor=east] (superc) at (-1,1.25)   {Superconducting};
  \node[anchor=east] (ion) at (-1,-0.25)   {Ion-Trap};
  
  \node[anchor=east] (8) at (2,2)   {8 Qubits};
  \node[anchor=east] (27) at (2,1.5)   {27 Qubits};
  \node[anchor=east] (80) at (2,1)   {80 Qubits};
  \node[anchor=east] (127) at (2,0.5)   {127 Qubits};
  \node[anchor=east] (11) at (2,-0.25)   {11 Qubits};

  \node[anchor=east] (qiskit) at (4.5,1.25)   {Qiskit};
  \node[anchor=east] (tket) at (4.5,0.25)   {TKET};

  \node[anchor=east] (O0) at (8,2.25)   {O0};
  \node[anchor=east] (O1) at (8,1.75)   {O1};
  \node[anchor=east] (O2) at (8,1.25)   {O2};
  \node[anchor=east] (O3) at (8,0.75)   {O3};
  
  \node[anchor=east] (line) at (8,0.25)   {Line Placement\strut};
  \node[anchor=east] (graph) at (8,-0.25)   {Graph Placement\strut};
  \draw[-latex] (start) -- (ion.west);
  \draw[-latex] (start) -- (superc.west);
  \draw[-latex] (ion.east) -- (11.west);
  \draw[-latex] (superc.east) -- (8.west);
  \draw[-latex] (superc.east) -- (27.west);
  \draw[-latex] (superc.east) -- (80.west);
  \draw[-latex] (superc.east) -- (127.west);
  
  \draw[-latex] (8.east) -- ([xshift=0mm, yshift=2mm]qiskit.west);
  \draw[-latex] (27.east) -- ([xshift=0mm, yshift=1mm]qiskit.west);
  \draw[-latex] (80.east) -- ([xshift=0mm, yshift=0]qiskit.west);
  \draw[-latex] (127.east) -- ([xshift=0mm, yshift=-1mm]qiskit.west);
  \draw[-latex] (11.east) -- ([xshift=0mm, yshift=-2.0mm]qiskit.west);
  
  \draw[-latex] (8.east) -- ([xshift=0mm, yshift=2mm]tket.west);
  \draw[-latex] (27.east) -- ([xshift=0mm, yshift=1mm]tket.west);
  \draw[-latex] (80.east) -- ([xshift=0mm, yshift=0mm]tket.west);
  \draw[-latex] (127.east) -- ([xshift=0mm, yshift=-1mm]tket.west);
  \draw[-latex] (11.east) -- ([xshift=0mm, yshift=-2mm]tket.west);
  
  \draw[-latex] (qiskit.east) -- (O0.west);
  \draw[-latex] (qiskit.east) -- (O1.west);
  \draw[-latex] (qiskit.east) -- (O2.west);
  \draw[-latex] (qiskit.east) -- (O3.west);
  \draw[-latex] (tket.east) -- (line.west);
  \draw[-latex] (tket.east) -- (graph.west);

\end{tikzpicture}
	\caption{Search space of compilation options for the setup described in \autoref{ex:compilation_paths}.}\vspace*{-1mm}
	\label{fig:search_space}
\end{figure*}

\begin{example}\label{ex:compilation_paths}
	For illustration purposes assume that in the following, the \mbox{end-user} has to decide between four devices based on superconducting qubits (with $8$, $27$, $80$, and $127$ qubits, respectively) and a single ion trap-based one (with $11$ qubits).
	Additionally, assume that IBM's Qiskit~\cite{qiskit} and Quantinuum's TKET~\cite{sivarajahKetRetargetableCompiler2020} are available as \mbox{state-of-the-art} representatives for compilers.
	Last but not least, assume that in the case of Qiskit four different optimization levels (called $O0$ to $O3$) are considered, while in the case of TKET two different qubit placement algorithms (called \emph{Line~Placement} and \emph{Graph~Placement}) can be used for the compilation.

	Both considered compilers are capable of compiling a quantum circuit for all the considered devices.
	Therefore, the corresponding search space for compilation options is structured as illustrated in \autoref{fig:search_space}---comprising a total of $30$ different combinations of compilation options.
\end{example}

\subsection{Evaluation Metric}
\label{sec:eval_metric}

Based on the constructed search space, the definition of an \emph{evaluation metric}---determining whether a combination of compilation options is good for a certain quantum circuit---is an essential part of the proposed framework.
All further steps aim to optimize the prediction quality of the resulting model according to \emph{this} evaluation metric.

In principle, this evaluation metric can be designed to be arbitrarily complex, e.g., factoring in actual costs of executing quantum circuits on the respective platform or availability limitations for certain devices.
However, any suitable evaluation metric should at least consider the characteristics of the compiled quantum circuit and the respective device.
Thus, hardware information for each of the devices (such as qubit coherence times and gate fidelities) needs to be gathered.
For some of these devices, this information might not be publicly available from the respective hardware vendors and, hence, needs to be estimated from comparable architectures, previous records, or insight knowledge.

\begin{example}\label{ex:eval_metric}
In the following, we employ an evaluation metric that considers two aspects:
\begin{enumerate}
\item The device chosen in the combination of compilation options has to have at least as many physical qubits as the compiled circuit (or else the circuit would not be executable at all).
If this is not the case, the \mbox{worst-possible} score is assigned to this combination of compilation options.
\item If the first criterion is satisfied, each combination of compilation options is assigned an evaluation score defined by the formula:
\[ 
\mathit{eval~score}=\prod_{i=1}^{|G|} \mathit{gate~fid}(g_i) \prod_{j=1}^{m} \mathit{readout~fid}(q_j),
\]
where $\mathit{gate~fid}(g_i)$ denotes the expected fidelity of gate~$g_i$ on its corresponding qubit(s), $\mathit{readout~fid}(q_j)$ denotes the expected fidelity of a measurement operation~$q_j$ on its corresponding qubit, $|G|$ denotes the number of gates in the compiled circuit and $m$ denotes the number of measurements.
\end{enumerate}
This evaluation metric measures the expected fidelity of the compiled circuit due to noise, i.e., the probability of the circuit working as expected---higher values meaning less noise and, hence, better results.
\end{example}

\subsection{Generation of Training Data}\label{sec:training_data}
As with any machine learning algorithm, the availability of sufficient training data is critical for the performance of the resulting model.
In this regard, the biggest challenge is to gather sufficiently many quantum circuits to start the generation of training circuits---even more so, ones that are representative for the diversity of quantum computing applications.

Once a sufficiently large set of circuits has been gathered, \emph{all} possible combinations of compilation options are executed and evaluated using the metric from \autoref{sec:eval_metric} for each of the training circuits using the pipeline from \autoref{sec:pipeline}.
Then, the best combination of compilation options is stored as the classification label for this particular training circuit.

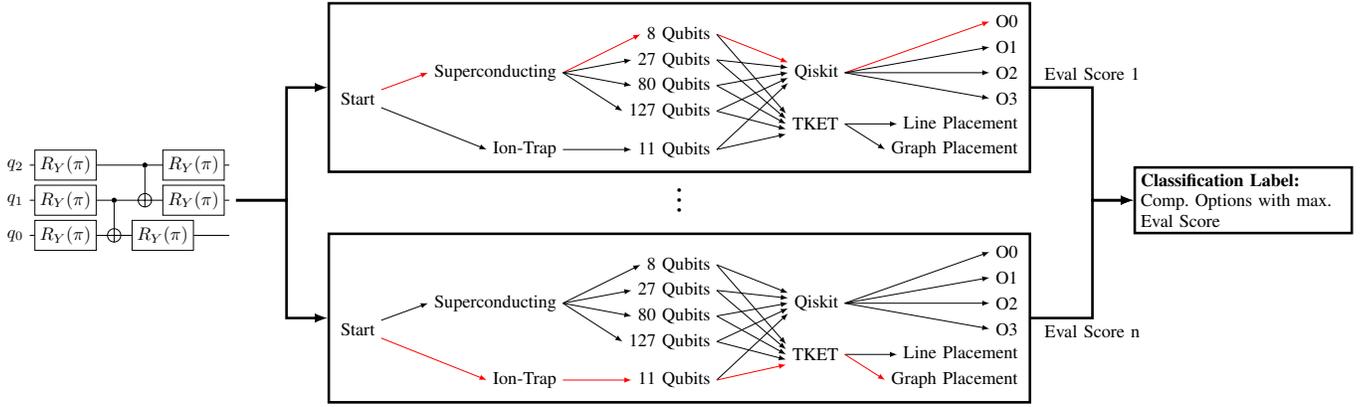
\begin{figure*}[t]
\centering
\resizebox{1.00\linewidth}{!}{
    \begin{tikzpicture}
    \node  (qc){
    \begin{tikzpicture}
      \begin{yquant}
            qubit {${q_2}$} q;    	
            qubit {${q_1}$} q[+1];
            qubit {${q_0}$} q[+1];
            
            box {$R_Y(\pi)$} q[0];
            box {$R_Y(\pi)$} q[1];
            box {$R_Y(\pi)$} q[2];
            
            cnot q[2] | q[1];
            cnot q[1] | q[0];
            
            box {$R_Y(\pi)$} q[0];
            box {$R_Y(\pi)$} q[1];
            box {$R_Y(\pi)$} q[2];
      \end{yquant}
    \end{tikzpicture}};
    \node [right of = qc, xshift=10cm, yshift=2.2cm, draw, line width=0.5mm] (top){
\begin{tikzpicture}
  \node (start) at (-5,0.75)   {Start};
  
  \node[anchor=east] (superc) at (-1,1.25)   {Superconducting};
  \node[anchor=east] (ion) at (-1,-0.25)   {Ion-Trap};
  
  \node[anchor=east] (8) at (2,2)   {8 Qubits};
  \node[anchor=east] (27) at (2,1.5)   {27 Qubits};
  \node[anchor=east] (80) at (2,1)   {80 Qubits};
  \node[anchor=east] (127) at (2,0.5)   {127 Qubits};
  \node[anchor=east] (11) at (2,-0.25)   {11 Qubits};

  \node[anchor=east] (qiskit) at (4.5,1.25)   {Qiskit};
  \node[anchor=east] (tket) at (4.5,0.25)   {TKET};

  \node[anchor=east] (O0) at (8,2.25)   {O0};
  \node[anchor=east] (O1) at (8,1.75)   {O1};
  \node[anchor=east] (O2) at (8,1.25)   {O2};
  \node[anchor=east] (O3) at (8,0.75)   {O3};
  
  \node[anchor=east] (line) at (8,0.25)   {Line Placement\strut};
  \node[anchor=east] (graph) at (8,-0.25)   {Graph Placement\strut};
  \draw[-latex] (start) -- (ion.west);
  \draw[-latex,red] (start) -- (superc.west);
  \draw[-latex] (ion.east) -- (11.west);
  \draw[-latex,red] (superc.east) -- (8.west);
  \draw[-latex] (superc.east) -- (27.west);
  \draw[-latex] (superc.east) -- (80.west);
  \draw[-latex] (superc.east) -- (127.west);
  
  \draw[-latex,red] (8.east) -- ([xshift=0mm, yshift=2mm]qiskit.west);
  \draw[-latex] (27.east) -- ([xshift=0mm, yshift=1mm]qiskit.west);
  \draw[-latex] (80.east) -- ([xshift=0mm, yshift=0]qiskit.west);
  \draw[-latex] (127.east) -- ([xshift=0mm, yshift=-1mm]qiskit.west);
  \draw[-latex] (11.east) -- ([xshift=0mm, yshift=-2.0mm]qiskit.west);
  
  \draw[-latex] (8.east) -- ([xshift=0mm, yshift=2mm]tket.west);
  \draw[-latex] (27.east) -- ([xshift=0mm, yshift=1mm]tket.west);
  \draw[-latex] (80.east) -- ([xshift=0mm, yshift=0mm]tket.west);
  \draw[-latex] (127.east) -- ([xshift=0mm, yshift=-1mm]tket.west);
  \draw[-latex] (11.east) -- ([xshift=0mm, yshift=-2mm]tket.west);
  
  \draw[-latex,red] (qiskit.east) -- (O0.west);
  \draw[-latex] (qiskit.east) -- (O1.west);
  \draw[-latex] (qiskit.east) -- (O2.west);
  \draw[-latex] (qiskit.east) -- (O3.west);
  \draw[-latex] (tket.east) -- (line.west);
  \draw[-latex] (tket.east) -- (graph.west);
  \end{tikzpicture}};
    
    \node [below of = top, yshift=-3.5cm, draw, line width=0.5mm] (bottom) {
    \begin{tikzpicture}
  \node (start) at (-5,0.75)   {Start};
  
  \node[anchor=east] (superc) at (-1,1.25)   {Superconducting};
  \node[anchor=east] (ion) at (-1,-0.25)   {Ion-Trap};
  
  \node[anchor=east] (8) at (2,2)   {8 Qubits};
  \node[anchor=east] (27) at (2,1.5)   {27 Qubits};
  \node[anchor=east] (80) at (2,1)   {80 Qubits};
  \node[anchor=east] (127) at (2,0.5)   {127 Qubits};
  \node[anchor=east] (11) at (2,-0.25)   {11 Qubits};

  \node[anchor=east] (qiskit) at (4.5,1.25)   {Qiskit};
  \node[anchor=east] (tket) at (4.5,0.25)   {TKET};

  \node[anchor=east] (O0) at (8,2.25)   {O0};
  \node[anchor=east] (O1) at (8,1.75)   {O1};
  \node[anchor=east] (O2) at (8,1.25)   {O2};
  \node[anchor=east] (O3) at (8,0.75)   {O3};
  
  \node[anchor=east] (line) at (8,0.25)   {Line Placement\strut};
  \node[anchor=east] (graph) at (8,-0.25)   {Graph Placement\strut};
  \draw[-latex,red] (start) -- (ion.west);
  \draw[-latex] (start) -- (superc.west);
  \draw[-latex, red] (ion.east) -- (11.west);
  \draw[-latex] (superc.east) -- (8.west);
  \draw[-latex] (superc.east) -- (27.west);
  \draw[-latex] (superc.east) -- (80.west);
  \draw[-latex] (superc.east) -- (127.west);
  
  \draw[-latex] (8.east) -- ([xshift=0mm, yshift=2mm]qiskit.west);
  \draw[-latex] (27.east) -- ([xshift=0mm, yshift=1mm]qiskit.west);
  \draw[-latex] (80.east) -- ([xshift=0mm, yshift=0]qiskit.west);
  \draw[-latex] (127.east) -- ([xshift=0mm, yshift=-1mm]qiskit.west);
  \draw[-latex] (11.east) -- ([xshift=0mm, yshift=-2.0mm]qiskit.west);
  
  \draw[-latex] (8.east) -- ([xshift=0mm, yshift=2mm]tket.west);
  \draw[-latex] (27.east) -- ([xshift=0mm, yshift=1mm]tket.west);
  \draw[-latex] (80.east) -- ([xshift=0mm, yshift=0mm]tket.west);
  \draw[-latex] (127.east) -- ([xshift=0mm, yshift=-1mm]tket.west);
  \draw[-latex,red] (11.east) -- ([xshift=0mm, yshift=-2mm]tket.west);
  
  \draw[-latex] (qiskit.east) -- (O0.west);
  \draw[-latex] (qiskit.east) -- (O1.west);
  \draw[-latex] (qiskit.east) -- (O2.west);
  \draw[-latex] (qiskit.east) -- (O3.west);
  \draw[-latex] (tket.east) -- (line.west);
  \draw[-latex,red] (tket.east) -- (graph.west);
  \end{tikzpicture}
    };
    \node [right of = qc, xshift=21cm, draw, text width=4cm, line width=0.5mm] (res)  {\textbf{Classification Label:} Comp. Options with max. Eval Score};
    
    \draw[-latex, line width=0.5mm] (qc.east)  -| ([xshift=-8mm, yshift=0mm]top.west) |- (top.west);
    \draw[-latex, line width=0.5mm] (qc.east)  -| ([xshift=-8mm, yshift=0mm]bottom.west) |- (bottom.west);

    \draw[-latex, line width=0.5mm] (top.east)  -| node[above] {Eval Score 1}([xshift=-8mm, yshift=0mm]res.west) |- (res.west);
    \draw[-latex, line width=0.5mm] (bottom.east)  -| node[below] {Eval Score n}([xshift=-8mm, yshift=0mm]res.west) |- (res.west);

    \node[align=center, rotate=90] at (11, 0) {\huge{$...$}};		
    \end{tikzpicture}}
    \caption{Generation of a training circuit from an example circuit.}
\label{fig:training_data}	
\end{figure*}

\begin{example}
	In order to generate training data based on the compilation options from~\autoref{ex:compilation_paths} and the evaluation metric from \autoref{ex:eval_metric}, the benchmark library MQT Bench~\cite{quetschlich2022mqtbench} has been used.
	It provides a large selection of benchmarks covering all kinds of quantum computing applications on various abstraction levels.
	Here, $3000$ benchmarks have been utilized from the \enquote{target-independent} level from $2$ to $127$ qubits.
	Using a timeout of $300$s for each combination of compilation options (as a \mbox{trade-off} between execution time and the number of successful compilations), classification labels have been determined as illustrated in \autoref{fig:training_data}.
	For each training circuit, all suitable combinations of compilation options are executed once to determine their corresponding evaluation scores. 
	Subsequently, the combination of compilation options leading to the highest evaluation score is the best combination of compilation options and thus the classification label for that training circuit. 
\end{example}

After generating the labeled training data---consisting of the initial quantum circuits and the best combinations of compilation options---the training circuits are transformed into \emph{feature vectors} to be suitable for training a classifier.
On the basis of that, the classifier is able to predict good combinations of compilation options for quantum circuits not used as training data (referred to as \emph{unseen} test circuits in the following) based on their features.
As for the evaluation metric, the features extracted from the input quantum circuit can be designed to be arbitrarily complex.

\begin{example}\label{ex:training_samples}
	After determining the correct classification label, the feature vector of each circuit is created.
	To this end, the number of qubits, the depth of the circuit, and the number of gates for each gate type according to the \emph{OpenQASM 2.0} specification~\cite{crossOpenQuantumAssembly2017} are used as features.
	Additionally, the following five composite features (adapted from \cite{supermarq}) are used:
	\begin{itemize}
	\item \emph{Program Communication}: Metric to measure the average degree of interaction for all qubits. A value of $1$ indicates that each qubit interacts at least once with all other qubits.
	\item \emph{Critical Depth}: Metric to measure how many of all \mbox{multi-qubit} gates are on the longest path (defining the depth of a quantum circuit).
	A value of $1$ indicates that all \mbox{multi-qubit} gates are on the longest path.
	\item \emph{Entanglement Ratio}: Metric to measure how many gates in a quantum circuit are \mbox{multi-qubit} gates. A value of $1$ indicates that the quantum circuit consists of only \mbox{multi-qubit} gates.
	\item \emph{Parallelism}: Metric to measure how much parallelization within the circuit is possible due to simultaneous gate execution. A value of $1$ describes large parallelization.
	\item \emph{Liveness}: Metric to measure how often the qubits are idling and waiting for their next gate execution. A value of $1$ describes a circuit in which there is a gate execution on each qubit at each time step.
\end{itemize}

\noindent In order to reduce the dimensionality of the resulting training circuits, the features that are zero for all training circuits are discarded.
	This resulted in $31$ features for each of the training circuits.
\end{example}

\subsection{Solving the Classification Task}

Eventually, the goal is to predict good combinations of compilation options for unseen test quantum circuits without executing and evaluating all possible combinations.
This classification task is perfectly suited for supervised machine learning classifiers.
To this end, a model representing the complex factors that influence the selection of compilation options can be trained without giving explicit guidelines.

\section{Resulting Prediction Framework \\and its Performance}
\label{sec:prob_desc}
The methodology proposed above has been implemented as a \mbox{proof-of-concept}, \mbox{open-source} Python package and is available on GitHub (\url{https://github.com/cda-tum/MQTPredictor}) as part of the Munich Quantum Toolkit (MQT).
It comprises three main functionalities:
\begin{enumerate}
\item Real-time prediction of good compilation options based on a \mbox{pre-trained} classifier,
\item running the compilation with the predicted best combination of compilation options itself, and,
\item the adaption and extension of the training pipeline to create custom prediction models.
\end{enumerate}
To this end, scikit-learn~\cite{scikit-learn} is used to train the underlying machine learning model.
The resulting framework allows for automatic prediction of good combinations of compilation options out of all considered possibilities. 
As a result, the tedious task of choosing a combination of compilation options is shifted from the \mbox{end-user} without quantum computing expertise to an automatic framework that has this domain knowledge embedded.
In the following, we describe the setup of the prediction framework and, then, comprehensively evaluate it in order to demonstrate the feasibility and effectiveness of the proposed methodology.

\subsection{Setting Up the Prediction Framework}
\label{sec:exp_setup}
In order to instantiate the framework as described above, it needs to be set up accordingly. To this end, the following steps need to be conducted:
\begin{enumerate}
	\item All qubit technologies, devices, compilers, and settings that should be considered need to be provided. Finally, this defines the set of possible combinations of compilation options.
	\item The corresponding SDKs and software packages need to be set up and stitched together.
	\item All information about the devices which shall be considered in the evaluation of the goodness of compilation options needs to be collected and, afterwards, incorporated into an evaluation metric.
	\item Finally, the classifier must be trained using a sufficient number of suitable training circuits. 
\end{enumerate}

\noindent Note that using this setup procedure, even future developments can easily be considered as it only requires the incorporation and/or adjustment of the respective instantiation.

In the following and for the purpose of evaluation, we considered the following setup (which was described throughout all previous examples):
\begin{itemize}
	\item Qubit Technology: superconducting and ion-trap-based qubits
	\item Devices: Archictures with $8$, $27$, $80$, and $127$ qubits for superconducting and $11$ qubits for the ion-trap based qubit technology
	\item Compiler: IBM's Qiskit (version $0.39.2$) and Quantinuum's TKET (version $1.9.0$)
	\item Compiler settings: $4$ optimization levels (Qiskit) and $2$ placement strategies (TKET)
\end{itemize}

\noindent Overall, this results in a total of $30$ different combinations of compilation options (as described in \autoref{ex:compilation_paths}).
As evaluation metric, we used the information and the corresponding calculation as previously described in \autoref{ex:eval_metric}. 
All circuits used for training are taken from the \mbox{MQT Bench} library~\cite{quetschlich2022mqtbench} (version $0.2.2$). 
More precisely, all circuits on the \enquote{target-independent} level from $2$ to $127$ qubits have been used---resulting in $3000$ training data samples.

Then, each of these circuits is exhaustively compiled with every possible combination of options and subsequently evaluated according to the desired evaluation metric.
Eventually, the best combination of options is recorded as the label for the training data\footnote{In many cases, it is beneficial to not discard the computed evaluation scores and the compiled circuits right away, but rather store them persistently. 
This allows one to validate the performance of a trained classifier, as well as quickly re-evaluate and re-label test samples after changes to the evaluation metric or the addition of new options.}.
Due to the independence of the individual generation jobs (regarding the evaluation metric as well as the combination of options), the training data generation can be easily parallelized across all available resources.
The generation of training data for this particular instantiation took around $100$h on a \mbox{$16$-threaded} Intel Xeon \mbox{W-$1370$P} with $3.60$ GHz and $128$ GB RAM, and resulted in a total of $38672$ compiled and evaluated circuits (covering all possible combinations of compilation options)---making the $3000$ training circuits ready for use in any machine learning algorithm.
In our evaluation, we used a $70\%$/$30\%$ \mbox{train-test-split} which resulted in $2100$ training samples and $900$ test samples. 
In the following, we summarize and discuss our evaluations on seven different supervised machine learning classifiers applied to the generated data.

\subsection{Resulting Performance}
\label{sec:exp_eval}
Using the instantiation described above, we evaluated the performance of various supervised machine learning classifiers---all instantiated and trained with \mbox{grid-searched} and \mbox{5-fold} \mbox{cross-validated} parameter values in order of minutes: 
\begin{itemize}
\item \emph{Random Forest} \cite{breiman2001random}
\item \emph{Gradient Boosting} \cite{friedman2001greedy}
\item \emph{Decision Tree} \cite{BreiFrieStonOlsh84}
\item \emph{Nearest Neighbor} \cite{cover1967nearest}
\item \emph{Multilayer Perceptron} \cite{haykin1994neural}
\item \emph{Support Vector Machine} \cite{cortes1995support}
\item \emph{Naive Bayes} \cite{bayes}
\end{itemize}
For an overview of today's use of those techniques, see~\cite{kotsiantis2007supervised}.

To this end, we considered the $900$ unseen test quantum circuits and predicted the best combination of compilation options for them.
To compare whether predictions actually yielded the best possible options (or not), the evaluation scores produced as part of the training data generation were persistently stored and used as a ground truth.
Subsequently, the resulting predictions have been ranked based on this ground truth.
That is, any prediction that actually constituted the best combination of compilation options got assigned rank~$1$, any prediction that constituted the second-best combination of compilation options got assigned rank~$2$, etc. With a total of $30$ combinations of compilation options, the worst prediction accordingly got assigned rank~$30$.

\begin{table}[t]
\caption{Comparison of different classifiers.}
\hspace{-0.6cm}
\resizebox{1.07\linewidth}{!}{
\begin{filecontents*}{performances.csv}
Classifier, Accuracy, Top3, Worst Rank, Eval. Score Diff., Std
Random Forest, 0.7692715756136183, 0.9556048834628191, 23, -0.0021, 0.0138 
Gradient Boosting, 0.756894016513969, 0.9456159822419534, 23, -0.0022, 0.0138 
Decision Tree, 0.7335889605248276, 0.9245283018867925, 12, -0.0038, 0.0202 
Nearest Neighbor, 0.7050605135165705, 0.9400665926748057, 23, -0.0036, 0.0219 
Multilayer Perceptron, 0.6612758737699355, 0.8890122086570478, 23, -0.0047, 0.0220 
Support Vector Machine, 0.620848320325755, 0.7569367369589345, 21, -0.0083, 0.0365 
Naive Bayes, 0.3320585906571655, 0.5482796892341842, 19, -0.0184, 0.0506 
\end{filecontents*}
\pgfplotstabletypeset[
create on use/mixed/.style={
    create col/assign/.code={%
        \edef\entry{\thisrow{Eval. Score Diff.}$\pm$\thisrow{Std}}
        \pgfkeyslet{/pgfplots/table/create col/next content}{\entry}
    }
},
col sep=comma,
	columns={Classifier, Accuracy, Top3, Worst Rank, mixed},
	columns/mixed/.style={column type={r},string type ,column name=Best Score Diff.},
     columns={Classifier, Accuracy, Top3, Worst Rank, mixed},
     columns/Classifier/.style={string type, column type={l}},
     columns/Accuracy/.style={column type={r}},
     columns/Top3/.style={column type={r}},
     columns/Worst Rank/.style={column type={r}},
     every head row/.style={before row=\toprule,after row=\midrule},
     every last row/.style={after row=\bottomrule},
]{performances.csv}
}

\label{tab:scores}
\end{table}

\begin{figure*}
\begin{center}
    \subfloat[Random Forest 
	\label{fig:RandomForestClassifier}]{
      \includegraphics[width=0.5001\textwidth]{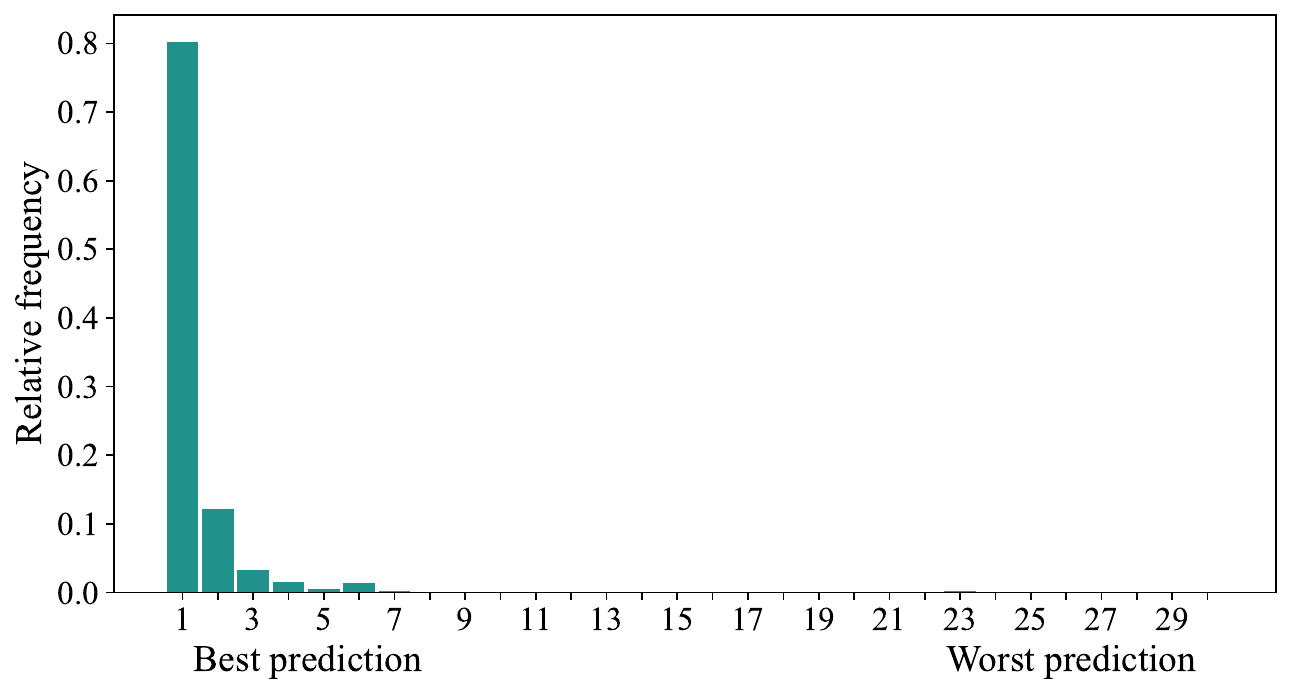}
    }
\end{center}

	\subfloat[Gradient Boosting
	\label{fig:GradientBoostingClassifier}]{%
      \includegraphics[width=0.5\textwidth]{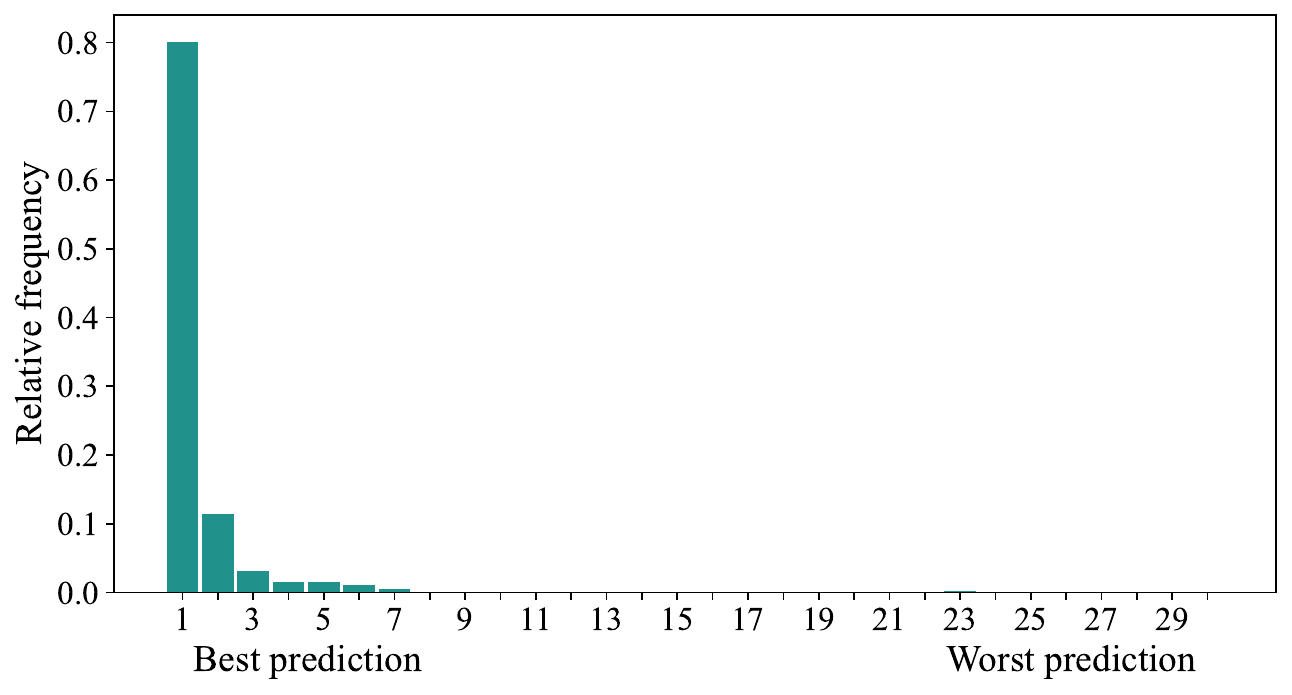}
    }
        \subfloat[Decision Tree 
	\label{fig:DecisionTreeClassifier}]{%
      \includegraphics[width=0.5\textwidth]{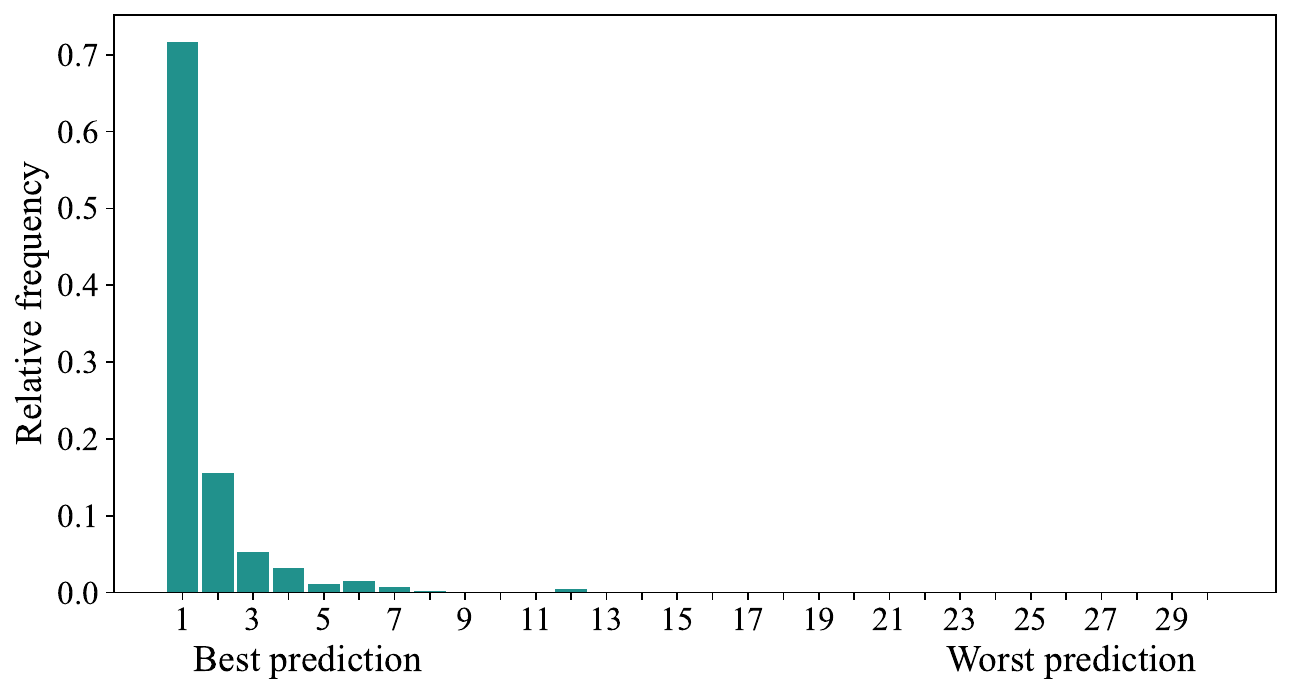}
    }
    
       \subfloat[Nearest Neighbor
	\label{fig:KNeighborsClassifier}]{%
      \includegraphics[width=0.5\textwidth]{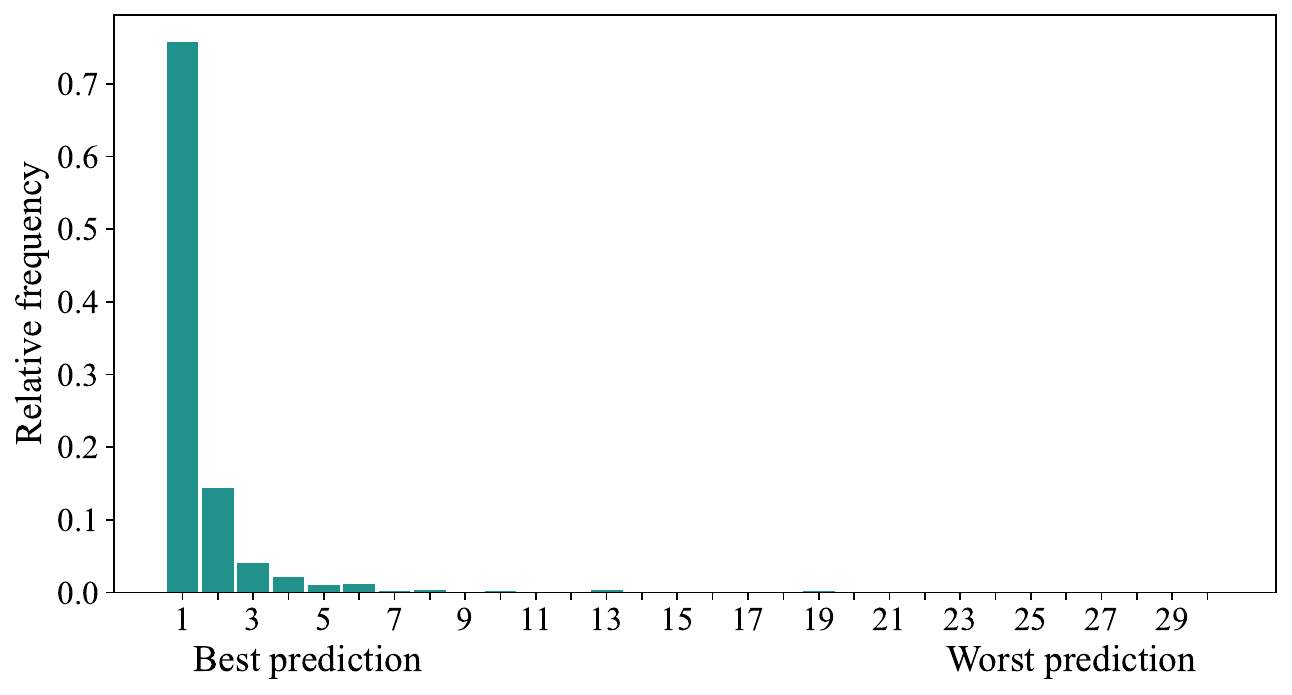}
    }
    \subfloat[Multilayer Perceptron
	\label{fig:MLPClassifier}]{%
      \includegraphics[width=0.5\textwidth]{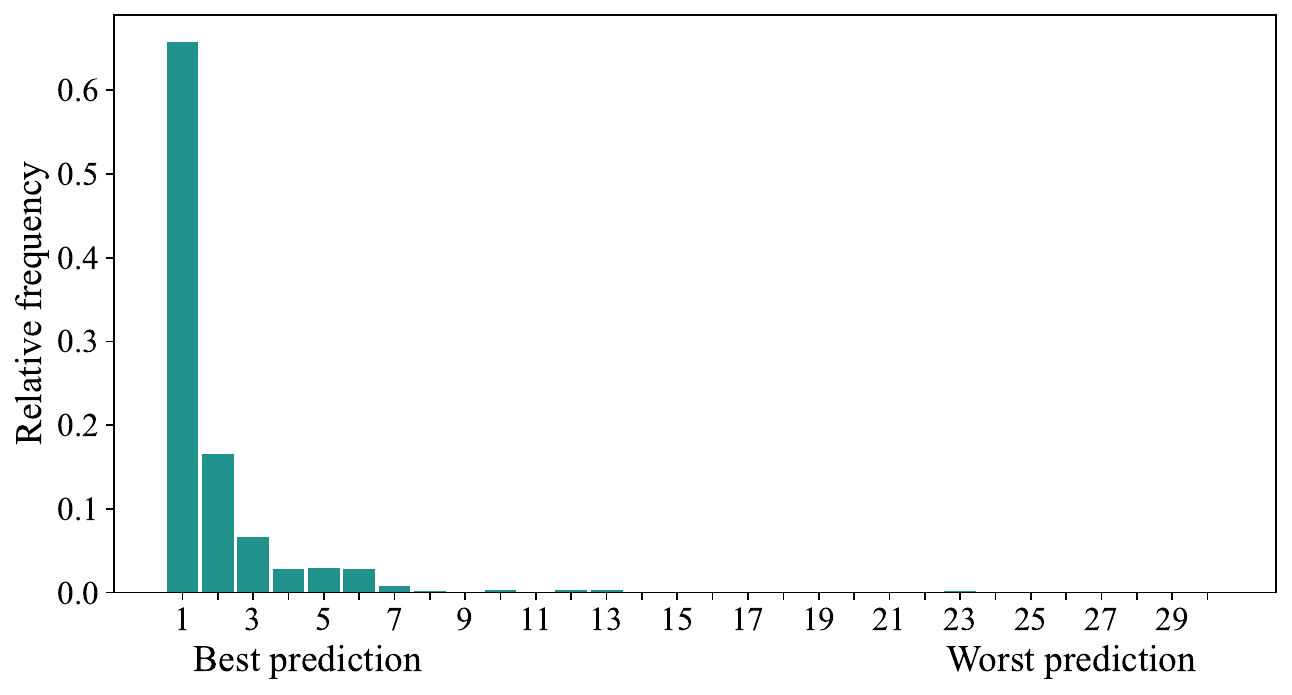}
    }

     \subfloat[Support Vector Machine
	\label{fig:SVM}]{%
      \includegraphics[width=0.5\textwidth]{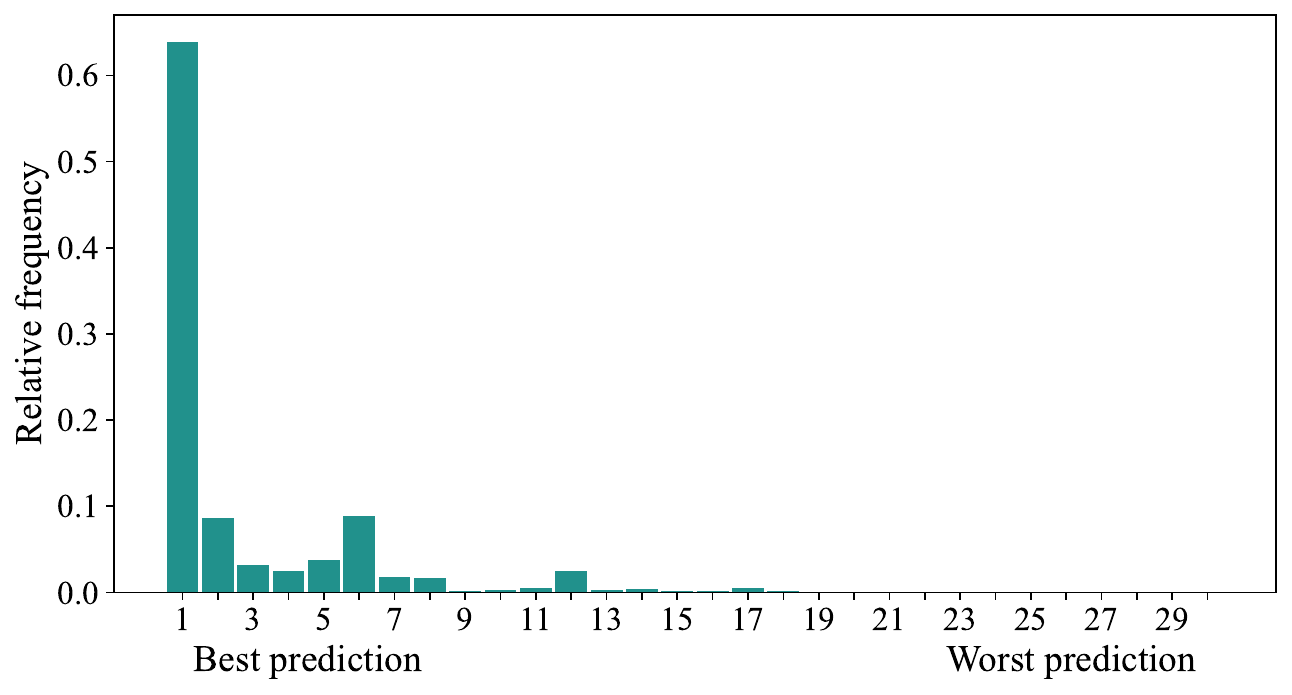}
    }
        \subfloat[Naive Bayes
	\label{fig:GaussianNB}]{%
      \includegraphics[width=0.5\textwidth]{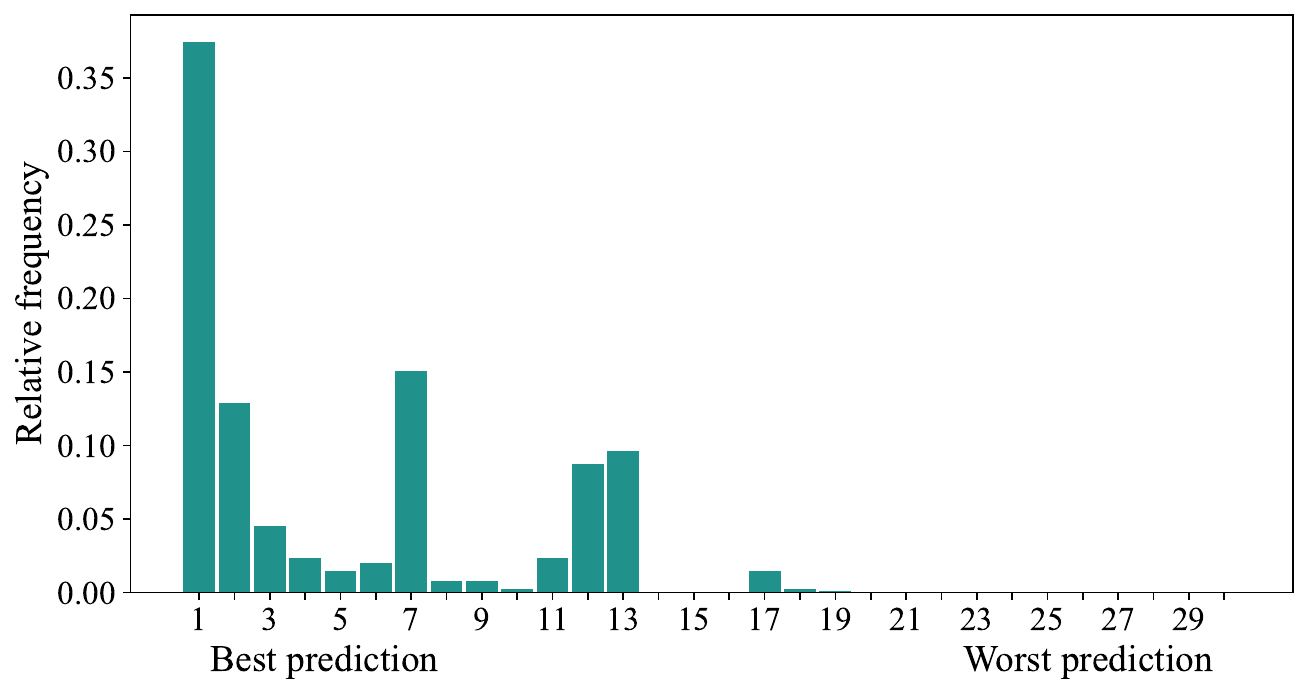}
    }

\caption{Comparison of supervised machine learning classifiers.}
\label{fig:eval_hist}
\end{figure*}

The results obtained correspondingly for all trained classifiers are summarized in \autoref{fig:eval_hist}. 
More precisely, the histograms show the relative frequency of the rankings for all the predictions made, ranging from the Random Forest classifier shown in \autoref{fig:RandomForestClassifier} to the Naive Bayes classifier shown in \autoref{fig:GaussianNB}.
The performance of all classifiers is assessed by four measures denoted in \autoref{tab:scores}:
\begin{enumerate}
\item Accuracy: Relative frequency of predicting the best possible combination of compilation options.
\item Top3: Relative frequency of predicting one of the \mbox{top-three} combinations of compilation options.
\item Worst Rank: Worst predicted rank for any of the test samples (out of $30$).
\item Best Score Diff.: Mean \emph{absolute} evaluation score difference (and standard deviation) compared to the performance of the best combination of compilation options.
\end{enumerate}

\noindent In this instantiation, the Random Forest classifier shown in \autoref{fig:RandomForestClassifier} with the following parameters led to the best performance:
\begin{itemize}
\item Number of Decision Trees: 100
\item Maximal depth = 26
\item Minimal samples per leaf = 2
\item Minimal samples per split = 2
\end{itemize}

\noindent It clearly confirms the quality of the predictions produced by the proposed prediction framework. 
In fact, for more than three quarters of all unseen test circuits, 
the \emph{best} combination of compilation options has been determined by the prediction framework. 
Moreover, for more than $95\%$ of the circuits, a combination of compilation options within the \mbox{top-three} is determined while the average absolute difference to the best performing options is around $0.2\%$ in expected fidelity.
Consequently, the Random Forest classifier yields the best or at least a very good prediction of the combination of compilation options in all test cases. 
Considering that all these predictions are made in \mbox{real-time} and must only be compiled once, this is a tremendous improvement compared to the state of the art where the end-users have to manually compile their circuit for all combination of compilation options to determine the best one---leading to a median runtime improvement of more than one order of magnitude.

\begin{figure*}[t]
	\begin{center}
		\includegraphics[width=0.99\linewidth]{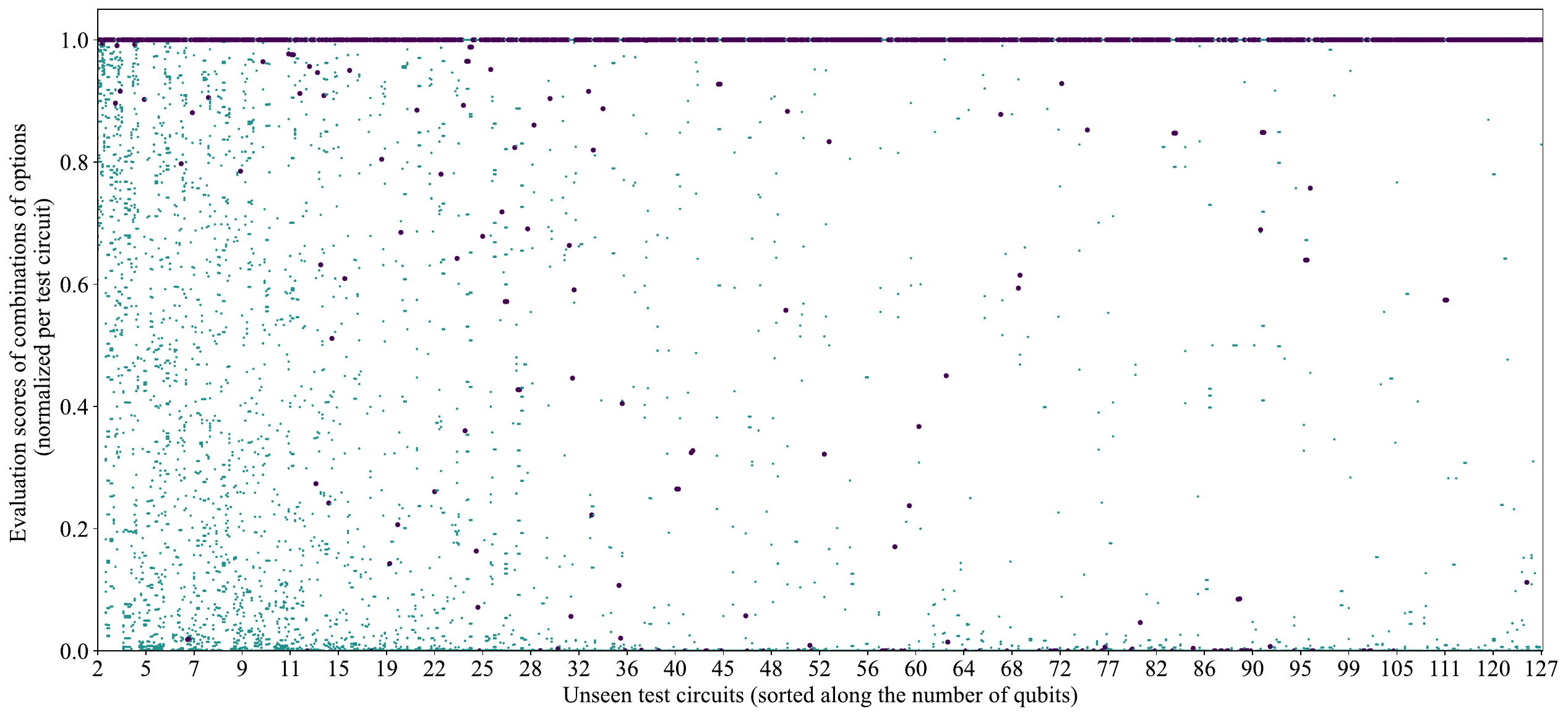}
	\end{center}
	\caption[Detailed evaluation of all combinations of compilation options for all unseen test circuits.]{Detailed evaluation of all combinations of compilation options for all unseen test circuits.
	All unseen test circuits are sorted by their qubit number which is indicated on the x-axis and, for most qubit numbers, multiple test circuits are evaluated. 
	For each test circuit, all executed and evaluated combinations of compilation options are indicated by green dots (\tikz{\draw[fill=color1,line width=0pt]  circle(0.7ex);}) while the predicted combination of compilation options is indicated by a purple dot (\tikz{\draw[fill=color2,line width=0pt]  circle(0.7ex);}).
	Due to the increased availability of quantum circuits with small numbers of qubits, the test circuits are not evenly spaced out with regard to their number of qubits.
	Since each device comes with a qubit limit (e.g., $8$, $11$, $27$, $80$, and $127$ qubits), the number of executed combinations of compilation options (and corresponding green dots) decreases with a growing number of qubits.
For the majority of the test cases, the predicted combination of compilation options returns the best results---illustrating the performance of the proposed framework.
	}
	\label{fig:dot_graph}
\end{figure*}

\begin{example}
Consider the case that a user wants to find the best combination of compilation options for a \mbox{\emph{Deutsch-Jozsa}} algorithm~\cite{deutsch1992rapid} instance with $7$ qubits (also taken from the \mbox{MQT Bench} library~\cite{quetschlich2022mqtbench}) with the setup described in \autoref{sec:exp_setup}. 
For that, the respective quantum circuit must be compiled for all possible combinations leading to $30$ compilations, where each resulting compiled circuit needs to be additionally evaluated using the evaluation metric described in \autoref{ex:eval_metric}---taking roughly a minute.
In comparison, using the proposed framework, only a prediction of the best combination of compilation options and the respective single compilation is necessary.
This is conducted in less than a second and, in this case, also leads to the best combination of compilation options while the calculation time is reduced by more than one order of magnitude. 
\label{ex:runtime_comp}
\end{example}

This shows the potential of machine learning-based optimization for quantum circuit compilation---similar to the enhancement these techniques brought to the domain of classical compiler optimization.
The Random Forest classifier is provided as a \mbox{pre-trained} model within our Python package.
Next, the performance of this classifier is examined to underline the importance of selecting good combinations of compilation options.

\subsection{Importance of Compilation Options}
A more detailed insight of the complete results of the Random Forest classifier is given in \autoref{fig:dot_graph}, which shows the evaluation scores of all possible combinations of compilation options for all unseen test circuits. 
Here, each green dot~(\tikz{\draw[fill=color1,line width=0pt] circle(0.7ex);}) corresponds to one possible combination of compilation options while the predicted result is indicated by a purple dot~(\tikz{\draw[fill=color2,line width=0pt]  circle(0.7ex);}).
The resulting evaluation score, i.e., the expected fidelity of the circuit execution, is normalized and plotted on the y-axis (higher is better), such that the normalized evaluation score of the best combination of compilation options per test circuit is assigned a value of $1.0$.

The results clearly demonstrate the significant impact of the chosen combination of compilation options on the expected performance/quality of the considered circuit. 
In other words: Whether a good or bad combination of compilation options is chosen frequently makes the difference between a reliable execution or one that does not work at all\footnote{Of course, the scores and ranking of combinations of compilation options highly depend on the chosen evaluation metric. 
However, the expected fidelity used in this evaluation has proven to be suitable for judging the expected performance of a quantum circuit. In addition to that, as discussed in \autoref{sec:exp_setup}, the setting used here is just a representative and can be adjusted to correspondingly reflect other setups.}.
Consequently, end-users may have a brilliant quantum circuit design, but its performance can still be spoiled by choosing a bad combination of compilation options. 
This emphasizes the importance of providing \mbox{end-users} with good predictions on the compilation options. 
The results summarized in \autoref{fig:dot_graph} again confirm that the proposed framework can deliver on that (in the vast majority of cases, the purple dot, i.e., the predicted combination of compilation options, is on the top of the spectrum).

\def\drawwith{0.5mm}
\begin{figure*}[t]
\centering
    \centering
    \resizebox{0.78\linewidth}{!}{
    \begin{tikzpicture}
        \node [circle, draw, minimum size=\circlesize, line width=\drawwith] (1) at (0,0)   {$|$Qubits$|$};
        \node [circle, draw, minimum size=\circlesize, below left = 1cm and 4cm of 1, line width=\drawwith] (2) {Ent. ratio};
        \node [circle, draw, minimum size=\circlesize, below right = 1cm and 4cm of 1, line width=\drawwith] (3) {Parallelism};
        \node [circle, draw, minimum size=\circlesize, below left = 1cm and 15mm of 2, line width=\drawwith] (4) {Parallelism};
        \node [circle, draw, minimum size=\circlesize, below right = 1cm and 15mm of 2, line width=\drawwith] (5)  {Liveness};
        \node [circle, draw, minimum size=\circlesize, below left = 1cm and 15mm of 3, line width=\drawwith] (6) {Ent. ratio};
        \node [circle, draw, minimum size=\circlesize, below right = 1cm and 15mm of 3, line width=\drawwith] (7)  {Crit. depth};

        \node [rectangle, draw, minimum width=2cm, below left = 10mm and \horizontalspace of 4, text width = 2cm, align=center, line width=\drawwith] (8) {Combination of options 4};        
        \node [rectangle, draw, minimum width=2cm, below right = 10mm and \horizontalspace of 4, text width = 2cm, align=center, line width=\drawwith] (9)  {Combination of options 2};        
        \node [rectangle, draw, minimum width=2cm, below left = 10mm and \horizontalspace of 5, text width = 2cm, align=center, red, text=black, line width=\drawwith] (10) {Combination of options 3};        
        \node [rectangle, draw, minimum width=2cm, below right = 10mm and \horizontalspace of 5, text width = 2cm, align=center, line width=\drawwith] (11) {Combination of options 4};        
        \node [rectangle, draw, minimum width=2cm, below left = 10mm and \horizontalspace of 6, text width = 2cm, align=center, line width=\drawwith] (12) {Combination of options 1};        
        \node [rectangle, draw, minimum width=2cm, below right = 10mm and \horizontalspace of 6, text width = 2cm, align=center, line width=\drawwith] (13) {Combination of options 3};        
        \node [rectangle, draw, minimum width=2cm, below left = 10mm and \horizontalspace of 7, text width = 2cm, align=center, line width=\drawwith] (14){Combination of options 8};        
        \node [rectangle, draw, minimum width=2cm, below right = 10mm and \horizontalspace of 7, text width = 2cm, align=center, line width=\drawwith] (15) {Combination of options 6};   

        \draw[red, line width=\drawwith] (1)--node[above, yshift=2mm] {$\leq 11$}(2);
        \draw[line width=\drawwith] (1)--node[above,  yshift=2mm] {$> 11$}(3);
        
        \draw[line width=\drawwith]  (2)--node[above,  xshift=-4mm] {$\leq 0.65$}(4);
        \draw[red, line width=\drawwith] (2)--node[above,  xshift=4mm] {$> 0.65$}(5);

        \draw[line width=\drawwith]  (3)--node[above,  xshift=-4mm] {$\leq 0.31$}(6);
        \draw[line width=\drawwith]  (3)--node[above,  xshift=4mm] {$> 0.31$}(7);

        \draw[line width=\drawwith]  (4)--node[above,  xshift=-5mm] {$\leq 0.23$}(8);
        \draw[line width=\drawwith]  (4)--node[above,  xshift=5mm] {$> 0.23$}(9);

        \draw[red, line width=\drawwith] (5)--node[above,  xshift=-5mm] {$\leq 0.45$}(10);
        \draw[line width=\drawwith]  (5)--node[above,  xshift=5mm] {$> 0.45$}(11);

        \draw[line width=\drawwith]  (6)--node[above,  xshift=-5mm] {$\leq 0.06$}(12);
        \draw[line width=\drawwith]  (6)--node[above,  xshift=5mm] {$> 0.06$}(13);

        \draw[line width=\drawwith]  (7)--node[above,  xshift=-5mm] {$\leq 0.82$}(14);
        \draw[line width=\drawwith]  (7)--node[above,  xshift=5mm] {$> 0.82$}(15);
        
        \draw[line width=\drawwith]  (0,1.5)--(1.north);
    \end{tikzpicture}}
\caption{Simplified decision tree classifier.}
\label{fig:tree}
\end{figure*}
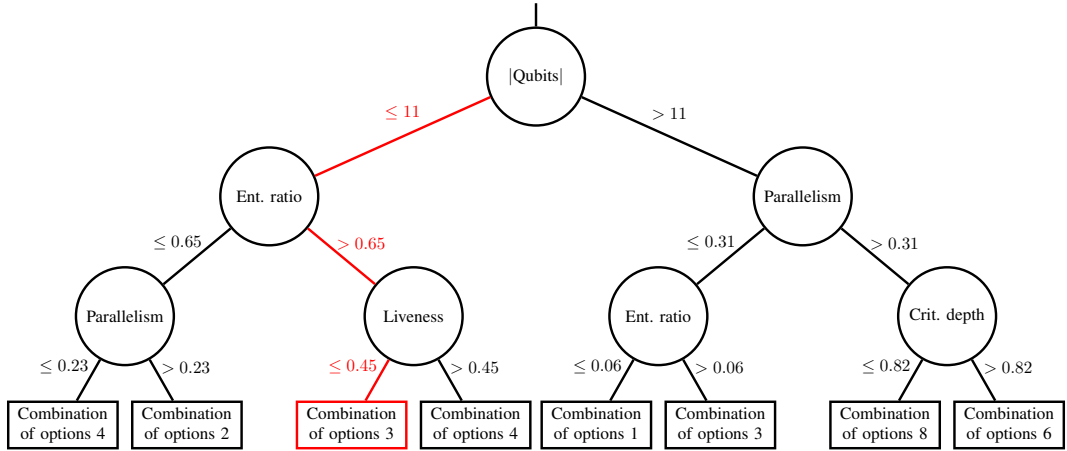

\section{Discussion}
\label{sec:discussion}

In general, machine learning techniques have two major drawbacks. 
The first is their \mbox{black-box} character and a lack of explainability of why a certain prediction has been made.
The second one is the effort spent on the model training and necessary preparations for that and, thus, the adaptability to change.
In this section, we discuss both drawbacks and propose approaches on how to mitigate and tackle the respective challenges.

\subsection{Knowledge Extraction}

Bringing light to opaque \mbox{black-box} machine learning techniques and learning more about their working mechanisms is a whole research area on its own (with overviews given in, e.g.,~\cite{burkart2021survey, e23010018}).
Depending on the machine learning classifier itself, the methods chosen to gain insight also vary. 

In the following, insights into the best performing classifier in the above evaluation, the Random Forest classifier, are given and discussed to extract explicit knowledge.
While it is particularly straightforward to get insights into this type of classifier, there are similar methods to extract information for other classifiers with overviews given in the provided references.
This can be used to guide further work towards exploring the potential of machine learning in predicting good combinations of compilation options and, additionally, to quickly verify the reasonableness of the trained classifier. 

The Random Forest classifier is composed of an ensemble of Decision Trees ($100$ in the evaluated scenario) that perform a majority vote for the prediction classification label and provides a rather simple method to gain insights called \emph{feature importance}.
The feature importance describes the normalized influence of each feature of the trained model and is defined as the reduction of the misclassification probability (also called \emph{gini impurity}) averaged over all comprised Decision Trees.

\begin{example}
\label{ex:decision_tree_example}
\autoref{fig:tree} shows a simplified illustration of a Decision Tree classifier for the scenario of determining good compilation options.
	Each node in such a Decision Tree corresponds to a decision depending on some of the features/characteristics of the circuit to be classified.
		A prediction follows a certain path from the root node according to the decisions at each node until a terminal node is reached---containing the predicted label/combination of compilation options.
	Using this classifier, for a circuit with $10$ qubits that has an entanglement ratio of $0.74$ and a liveness of $0.32$, \enquote{Combination~of~options~3} (highlighted in red) would be predicted as its classification label.
	For some circuits, the resulting prediction might not be the correct and best classification label and, hence, they are misclassified based on the decision nodes and their thresholds.
	With an ensemble of $100$ different Decision Trees, the influence of each feature on these misclassification rates can be determined and, by that, the reduction of the gini impurity.
\end{example}

\begin{figure}[t]
	\centering
	\includegraphics[width=0.99\linewidth]{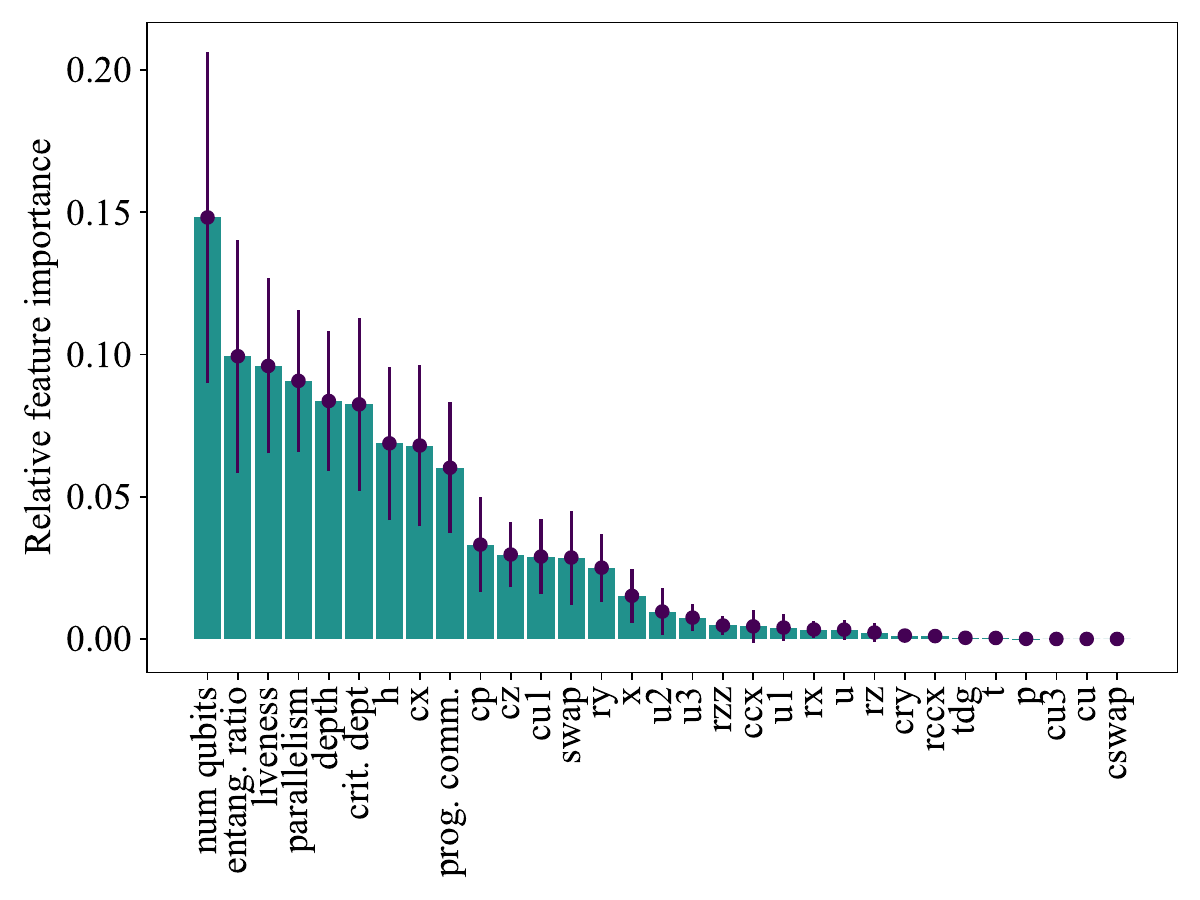}
	\caption{Feature importance: Normalized mean reduction of the gini impurity including its standard deviation for each feature.}
	\label{fig:feature_importances}
\end{figure}

The feature importance for the trained Random Forest classifier is summarized and denoted in decreasing order in \autoref{fig:feature_importances} with significant differences in the influence each feature has.
While $14$ of the $31$ features have a considerable influence on the trained model (all features up to the \emph{ry} gate count with a feature importance of $\geq2.5\%$), the remaining $17$ features are mostly negligible (with a feature importance of $<2.5\%$).
When focusing on the influential ones, the number of qubits feature is standing out with the highest importance. 
This is a strong testimony of the successful training of the underlying model and can be explained since no execution is possible at all if the device's number of qubits is smaller than needed.
Furthermore, the rather complex features (such as those adapted from~\cite{supermarq}) are considerably more influential than most of the plain gate counts. 
This indicates that those features characterize a quantum circuit better---with a few exemptions: the \emph{h}, \emph{cx}, \emph{cp}, \emph{cz}, \emph{cu1}, \emph{swap}, and \emph{ry} gates.

Since the \mbox{multi-qubit} gates (\emph{cx}, \emph{cp}, \emph{cz}, \emph{cu1}, and \emph{swap}) heavily influence the overhead during mapping, their high influence is expected.
Additionally, the \emph{cx}, \emph{cz}, and \emph{ry} gates are native to some devices, resulting in significantly fewer native gates after compilation.
Similarly, the \emph{h} gate is more efficiently compiled to certain native \mbox{gate-sets} leading to higher influence.

While it is always possible to assume the reason behind some feature's impact and deduct rules of thumb (such as \enquote{\emph{quantum circuits with a high program communication should be mapped to \mbox{ion-trap} devices due to their full connectivity}} or \enquote{\emph{circuits with a large \emph{cx} gate count should be mapped to devices where it is a native gate}}), it is far too complex to factor in and add weight to a large number of features---even with only $30$ different combinations of compilation options and $31$ features.
This underlines the importance and potential of machine learning approaches that aim for the same success they brought to the domain of classical compiler optimization.

Nevertheless, the extracted knowledge in the form of feature importance is helpful in two ways:
\begin{itemize}
\item It indicates what kind of features are helpful to characterize a quantum circuit and underlines the importance of thorough feature engineering.
\item It provides insight into the trained model, allowing one to quickly verify whether it is reasonable at all. 
\end{itemize}

\subsection{Adaptability to Change}
The second drawback of machine learning algorithms is the effort needed for the generation of training data and the model training itself---especially, since the model has to be \mbox{re-trained} whenever the evaluation metric or the available options are updated.

Quantum computers are calibrated frequently to ensure that they operate at their lowest error rate.
Thus, the noise characteristics considered in the proposed evaluation metric could change from calibration to calibration.

In addition to the characteristics of existing devices that change over time, the number of devices and their underlying technologies, compilers, and respective options is steadily increasing.
Naturally, adding more options increases the time to generate training data, while also potentially requiring an update to the evaluation scores and the classification label, since a new option might lead to better results than were possible before.

It is neither feasible nor realistic to always \mbox{re-generate} all the training data and \mbox{re-train} the model from scratch.
Calibration might be performed daily, while the generation of the training data on our system took roughly five days (while the training itself takes only minutes).
Obviously, large \emph{High Performance Computing (HPC)} systems could be used to speed up generation and training, but the amount of required resources might not be worth the price.

Another approach is to \mbox{re-use} the training data of the previously trained models as much as possible.
As described in \autoref{sec:exp_setup}, all compiled circuits used for the training data generation are persistently stored in the proposed framework.
Thus, all these compiled circuits can be utilized whenever new calibration data is available, since they must only be \mbox{re-evaluated} and not \mbox{re-generated}.
The updated calibration data just assigns a new value to each compiled circuit to determine the classification label.
Similarly, the same previous training can be used, even without any \mbox{re-evaluation}, whenever a new combination of compilation options is added---only the newly added compilation options must be applied to compile each training circuit before the classification label is determined.

Nevertheless, in both cases, the model itself must be \mbox{re-trained} based on adjusted training data.
There are different approaches to avoid the necessity of \mbox{re-training} the model from scratch.
\mbox{So-called} \emph{warmstart} approaches can be utilized where parameters of an existing model are used as the starting point for a new model (incorporating, for example, the latest calibration data) as, e.g., discussed in~\cite{10.5555/3495724.3496051} for neural networks.

\section{Conclusion}\label{sec:conclusion}
In this work, we proposed a methodology that allows \mbox{end-users} from domains different to quantum computing to realize their applications more easily on actual hardware.
This is accomplished by shifting the tedious selection of good compilation options away from the end-user and embedding the necessary expert knowledge in a prediction framework that takes a quantum circuit as input and predicts the best combination of various qubit technologies, devices, compilers, and corresponding settings.
Experimental evaluations show that, out of $30$ different combinations of compilation options, the most powerful classifier is capable of predicting the best combination of compilation options for more than three quarters of all unseen test circuits.
For more than $95\%$ of the circuits, a combination of compilation options within the \mbox{top-three} is determined---while the median compilation time is reduced by more than one order of magnitude compared to manually compiling for all possible combinations of compilation options and choosing the best result.
In addition to the time savings, the underlying model provides insight on why certain decisions have been made---allowing \mbox{end-users} to build up expertise from the predicted results.
Furthermore, the proposed methodology can easily be adapted and extended to future qubit technologies, devices, compilers, and compiler settings.
The corresponding framework and the \mbox{pre-trained} classifier are publicly available on GitHub (\url{https://github.com/cda-tum/MQTPredictor}) as part of the Munich Quantum Toolkit (MQT).
To the best of our knowledge, this work constitutes the first step towards using machine learning for quantum compilation---aiming for a similar success as achieved in classical compilation leading and, by that, simplifying and accelerating the adoption of quantum computing to solve problems from various application domains.

\section{Acknowledgements}
This work received funding from the European Research Council (ERC) under the European Union’s Horizon 2020 research and innovation program (grant agreement No. 101001318), was part of the Munich Quantum Valley, which is supported by the Bavarian state government with funds from the Hightech Agenda Bayern Plus, and has been supported by the BMWK on the basis of a decision by the German Bundestag through project QuaST, as well as by the BMK, BMDW, and the State of Upper Austria in the frame of the COMET program (managed by the FFG).

\vspace{5cm}

\clearpage
\printbibliography

\end{document}